\documentclass[twocolumn,showpacs,preprintnumbers,superscriptaddress,
               amsmath,amssymb,prd]{revtex4}

\usepackage{graphicx}
\usepackage{dcolumn}
\usepackage{bm}
\usepackage[dvips]{color}

\newcommand{\De}{\Delta}

\renewcommand{\>}{\rangle} 
\newcommand{\txt}{\textstyle}
\newcommand{\dsp}{\displaystyle}

\newcommand{\beq}{\begin{equation}}
\newcommand{\eeq}{\end{equation}}
\newcommand{\ba}{\begin{array}}
\newcommand{\ea}{\end{array}}
\newcommand{\bea}{\begin{eqnarray}}
\newcommand{\eea}{\end{eqnarray}}
\newcommand{\bi}{\begin{itemize}}  
\newcommand{\ei}{\end{itemize}}
\newcommand{\ben}{\begin{enumerate}} 
\newcommand{\een}{\end{enumerate}}

\newcommand{\half} {{\txt \frac{1}{2}}}
\newcommand{\third}{{\txt \frac{1}{3}}}

\newcommand{\threehalves}{{\txt \frac{3}{2}}}

\newcommand\hide[1]{}

\newcommand{\feyn}[1]{
  \setbox0=\hbox{\ensuremath{#1}}
  \hbox to\wd0{\hbox to0pt{\hbox to\wd0{\hss/\hss}\hss}\box0}}

\newcommand{\MeV}{{\rm MeV}}

\begin{document}

\preprint{MIT-CTP-3558}
\preprint{UMPP 05-022}

\title{A Hot Water Bottle for Aging Neutron Stars}

\author{Mark Alford}%
\affiliation{Physics Department, Washington University,
St.~Louis, MO~63130, USA}
\author{Pooja Jotwani}
\affiliation{Charles W. Flanagan High School, Pembroke Pines, FL 33029, USA}
\author{Chris Kouvaris}
\affiliation{Center for Theoretical Physics, Massachusetts
Institute of Technology, Cambridge, MA 02139, USA}
\author{Joydip Kundu}
\affiliation{Department of Physics, University of Maryland,
College Park, MD 20742, USA}
\author{Krishna Rajagopal}
\affiliation{Center for Theoretical Physics, Massachusetts
Institute of Technology, Cambridge, MA 02139, USA}

\date{November, 2004}

\begin{abstract}
The gapless color-flavor locked (gCFL)
phase is the second-densest phase of matter in
the QCD phase diagram, making it a plausible constituent
of the core of neutron stars.
We show that even a relatively small region of gCFL matter in a star
will dominate both the heat capacity $C_V$ and the heat loss by
neutrino emission $L_\nu$.
The gCFL phase is characterized by an unusual quasiparticle
dispersion relation that makes both
its specific heat $c_V$ and its neutrino emissivity $\varepsilon_\nu$ 
parametrically 
larger than in any other phase of nuclear or
quark matter. 
During the epoch
in which the cooling of the star is dominated by 
direct Urca neutrino emission, the presence of a gCFL region
does not strongly alter the cooling history because the
enhancements of $C_V$ and $L_\nu$ cancel against each other.
At late times, however, the cooling is dominated by
photon emission from the surface, so
$L_\nu$ is irrelevant, and the anomalously large heat capacity of
the gCFL region keeps the star warm. The temperature
drops with time as $T\sim t^{-1.4}$
rather than the canonical $T\sim t^{-5}$.  This provides
a unique and potentially 
observable signature of gCFL quark matter.
\end{abstract}

\pacs{Valid PACS appear here}

\maketitle

\section{Introduction}
\label{sec:intro}

It has often been suggested that the core of a neutron star
may contain quark matter in one of the color-superconducting
phases~\cite{Reviews}. The densest predicted phase 
on the QCD phase diagram is the
color-flavor locked (CFL) phase, which is a color superconductor but
an electromagnetic insulator~\cite{Alford:1998mk,Reviews}. 
The second-densest phase
is the gapless CFL (gCFL) phase, which is
a conductor with a nonzero density of 
electrons~\cite{Alford:2003fq,Alford:2004hz,Ruster:2004eg,Fukushima:2004zq}. 
It also has gapless quark quasiparticles,
one of which has an almost-quadratic
dispersion relation, arising without fine-tuning
because it is enforced by the
requirement that the matter be electrically 
neutral~\cite{Alford:2003fq,Alford:2004hz}.
We show in this paper that this characteristic feature
of the gCFL phase means that if quark matter in this
phase is present in a neutron star, it dominates
the heat capacity and neutrino luminosity, and therefore controls the 
cooling of the star.  At late times this produces a
unique signature, as the large heat capacity
keeps the star anomalously warm. A neutron star that is
tens of millions of years old
will be an order of magnitude or more warmer
if it contains a region of gCFL quark matter than if it does not.  

In any speculation about the phases of matter that occur
inside a neutron star, the main challenge is to provide
observable signatures of the presence of these phases.
Since we are proposing such a signature,
albeit one that presents significant observational
challenges, we first set the stage with a quick survey of previous
proposals.
\begin{list}{$\bullet$~} {\usecounter{enumi}
  \setlength{\topsep}{1ex} 
  \setlength{\itemsep}{-0.5\parsep} 
  \setlength{\labelwidth}{0em}
  \setlength{\labelsep}{0em}
  \setlength{\leftmargin}{0em} 
 }
\item Mass-radius relation.
If we could measure the mass and radius of several neutron stars
to a reasonable accuracy, mapping out the mass-radius
relationship, we would have a strong constraint on the
equation of state of dense matter. However, although 
such measurements would dramatically reduce the
current uncertainties in our knowledge of the
equation of state of the nuclear matter ``mantle''
of neutron stars, and could yield
evidence of some sort of exotic phase in the core, they would not
provide specific evidence of the presence of quark matter
\cite{Lattimer:2004pg,Alford:2004pf}.

\item Double pulsar timing. 
There is a good prospect that the long term analysis
of the recently discovered binary double pulsar~\cite{DoublePulsar} may yield
a measurement of the moment of inertia of a neutron star~\cite{Kramer:2004gj}.
This would provide information about the density profile
that is complementary to that obtained from a mass-radius
relation, as it would constrain the ``compactness'' of a star.

\item Gravitational waves from collisions.
If we could detect gravity waves from neutron stars spiraling into
black holes in binary systems, we could perhaps analyze them for
information about the density profile of the neutron star,
in particular the presence of an interface separating a denser
quark core from a less-dense nuclear mantle~\cite{Alford:2001zr}.

\item Spinning out a quark matter core.
If conditions are ``just so'',
rapidly spinning oblate neutron stars
may not have quark matter in their
cores even though more slowly rotating spherical neutron stars do. 
This could
be detected either by anomalies in braking indices of
stars that are ``spinning down''~\cite{Glendenning:1997fy} or by 
anomalous population statistics of stars that are being
``spun up'' by accretion~\cite{Glendenning:2000zz}.  Recent observations show
no sign of such an effect in the histogram of spin-frequencies
of stars in the act of being spun up~\cite{Chakrabarty:2004tp}, 
indicating that
if quark matter is present, spinning the star and making
it oblate does not get rid of it.
If there is a quark matter core, it must therefore occupy a reasonable
fraction of the star.

\item $r$-modes.
A rapidly spinning neutron star
will quickly slow down if it is unstable with respect to  
bulk flows known as $r$-modes, which transfer the star's
angular momentum into gravitational radiation.
This phenomenon will only occur if damping is sufficiently small,
so it provides a probe of the viscosity of the interior
of the star. Such arguments have been used to 
rule out
the possibility that pulsars are made entirely of CFL quark
matter~\cite{Madsen:1999ci}, 
in which viscous damping is negligible~\cite{Madsen:1999ci,Manuel:2004iv},
but their implications for the possibility of CFL quark
matter localized within the core of a neutron star
have not yet been analyzed.  Since gCFL quark matter is expected
to have a large viscosity, its
presence is unlikely to be constrained by $r$-mode arguments.

\item Core glitches. If the third-densest phase on the QCD phase
diagram is not nuclear matter, it must be a form of quark matter
with less pairing than in the gCFL phase.  A leading
candidate is the crystalline color superconducting
phase~\cite{Alford:2000ze}.  If this form of quark matter occurs
within the core of a neutron star, because it is both superfluid and
crystalline it may be a locus for pulsar glitches.
This proposal has not yet been worked out sufficiently
quantitatively to determine whether such core glitches
are observable and if so whether they
are consistent with (some) observed glitches.

\item Direct neutrino detection.
Neutrinos have a long mean free path even in nuclear matter,
so they can potentially carry information about the core
directly to the outside world. Not coincidentally, neutrinos
are very hard to detect, and the only time when a neutron star
emits enough neutrinos to be detectable on earth
is during the first few seconds after the supernova explosion.
The time-of-arrival distribution
of supernova neutrinos could teach us about possible phase
transitions to and in quark matter~\cite{Carter:2000xf,Jaikumar:2002vg}, 
but analysis of this
proposal requires a better understanding of both the
supernova itself and of the properties of quark matter at 
MeV temperatures, where the phase diagram of QCD is more
baroque than at zero temperature~\cite{Ruster:2004eg,Fukushima:2004zq}.

\item Cooling.
A much better prospect is the indirect detection of neutrino emission,
which is the dominant heat loss mechanism for the first million
years or so, and can therefore be inferred from measurements of
neutron star temperature as a function of age.
Moreover, because both neutrino emission rates
and heat capacity generally rise with density, neutron
star cooling is likely to be
preferentially sensitive to the properties of matter
in the core of a neutron star. 

The qualitative distinction among cooling behaviors 
that may be discerned from the
measurement of temperatures of stars that
are $10^{3-6}$ years old is between 
stars in which direct Urca processes are allowed (which 
yields a neutrino emissivity $\varepsilon_\nu \sim T^6$), 
leading to rapid cooling, and stars in which
direct Urca processes are 
forbidden~\cite{Pethick,ShapiroTeukolsky,Page:2000wt,Page:2004fy,Lattimer:2004pg,Yakovlev:2004yr}.  
Ordinary nuclear matter
is unusual in that its direct Urca processes
$n\rightarrow p+e+\bar \nu$ and $p+e\rightarrow n+\nu$
are kinematically forbidden, meaning that
neutrino emission relies 
upon slower reactions ($\varepsilon_\nu \sim T^8$).
Direct Urca processes 
are allowed in sufficiently dense nuclear matter, nuclear
matter with nonzero hyperon density~\cite{Hyperon}, pion 
condensation~\cite{BahcallWolf} or kaon condensation~\cite{Kaplan:1986yq}, 
and in all proposed phases of quark matter except CFL~\cite{Iwamoto}.
In the CFL phase, there are no direct Urca
processes because thermally excited quark quasiparticles
are exponentially rare. There are neutrino
emission processes involving
collective excitations that lead to 
$\varepsilon_\nu\sim T^{15}$~\cite{Jaikumar:2002vg}, but in reality any
CFL quark matter within a star will cool by conduction, not
by neutrino emission~\cite{Shovkovy:2002kv}.  
Indeed, because all forms of dense
matter are good heat conductors
the cooling of a star tends to be dominated by whichever phase
has the highest neutrino emissivity.  Hence, the discovery of fast cooling
would only tell us that some part of the star consists of one of the
many phases that allow direct Urca. Discovery of stars that cool
slowly would be an indication that they contain only
medium-density nuclear matter and perhaps CFL quark matter.


\end{list}

To date, none of the schemes listed above has provided an unambiguous
signature of the presence of quark matter, although all are the
subject of ongoing observational effort, which in turn drives
improvements on the theoretical side.
In this paper, we argue that recent theoretical advances
in our understanding of the properties of quark matter
offer the prospect of an unambiguous detection, if it
is possible to measure the temperatures of neutron
stars that are old enough that their cooling is 
no longer dominated by neutrino emission.
Admittedly, this presents an observational challenge.
However, it is a challenge that has not been closely studied
prior to our work, since all forms of dense matter
{\it except} gCFL quark matter result in neutron stars that
cool comparably (and very) rapidly in their old age.
We show that quark matter
in the gCFL phase keeps 
aged neutron stars (those significantly older than a million years)
much warmer than is predicted by any other assumed dense matter
physics. 

Younger neutron stars containing gCFL quark matter
have a faster-than-standard direct Urca
neutrino emissivity
$\varepsilon_\nu\sim T^{5.5}$, but this does
not lead to faster-than-standard-direct-Urca cooling
because of the correspondingly enhanced gCFL specific heat.

In Section \ref{sec:gCFL}, we introduce the relevant properties of
the gapless CFL phase of quark matter, and in Sections 
\ref{sec:specific_heat}
and \ref{sec:neutrino_emissivity}
we present the calculations of its specific
heat $c_V$ and neutrino emissivity $\varepsilon_\nu$, respectively.
These are our central calculational results.  We do not
provide a state-of-the-art calculation of the cooling
of a neutron star containing gCFL quark matter. Instead,
in Section \ref{sec:implications}, we provide an introduction to the physics
of neutron star cooling that suffices to illustrate
the qualitative consequences of the quantitative
results for the gCFL $c_V$ and $\varepsilon_\nu$.
The astrophysically inclined 
reader interested in our results  and their implications 
but not in their derivation can find $c_V$ in 
Eq.~(\ref{gCFLspecificheat})
and $\varepsilon_\nu$ in (\ref{emissivityresult}) and,
more conveniently, 
Fig.~\ref{fig:plot2}
and should read the text around these results and then
turn to Section \ref{sec:implications}.

\section{Introduction to the Gapless CFL Phase of Quark Matter}
\label{sec:gCFL}

At any densities that are high enough that nucleons are
crushed into quark matter, 
the quark matter that results at sufficiently low temperatures
is expected to be in one of a family of color superconducting
phases, with Cooper pairing of quarks
near their Fermi surfaces~\cite{Reviews}.
The QCD quark-quark interaction is strong and is attractive between
quarks that are antisymmetric in color.
If there is quark matter in the cores of neutron stars, we therefore
expect it to be color superconducting. 
The phenomenon persists to asymptotically high densities, where
the interaction becomes weak and {\it ab initio} calculations
of properties of color superconducting matter become 
rigorous~\cite{Reviews}.
The QCD phase diagram exhibits a rich structure of
color superconducting phases as a function
of temperature and density~\cite{Reviews,Ruster:2004eg,Fukushima:2004zq}, 
but in this paper we can simplify it
by working at zero temperature. This is reasonable because
we will be discussing neutron stars with temperatures in the keV
range, which is orders of magnitude
colder than the various critical temperatures at which
phase transitions between different quark matter phases occur.

\subsection{The CFL phase under stress}

At asymptotically high densities, 
where the up, down and strange quarks can be treated
on an equal footing and the disruptive effects of the
strange quark mass can be neglected, quark matter
is in the color-flavor locked (CFL) phase, in which
quarks of all three colors and all three flavors form
Cooper pairs~\cite{Alford:1998mk}.  The CFL phase is a color superconductor
but is an electromagnetic insulator, with zero electron density. 
In real-world quark matter,
as may exist in the cores of compact stars, the density is
not asymptotically high. The quark chemical potential $\mu$ is
of order 500~MeV at most, making it important to include
the effects of the strange quark mass $M_s$,
which is expected to be density dependent,
lying somewhere between the current mass $\sim 100$~MeV and the
vacuum constituent quark mass $\sim 500$~MeV.
To describe macroscopic regions of quark matter, we must 
also impose
electromagnetic and color
neutrality~\cite{Alford:2002kj,Neutrality,Gerhold:2004sk} 
and allow for equilibration under the weak
interactions. 
The CFL pairing pattern 
is antisymmetric in flavor, color, and spin,
so it involves pairing between different flavors. 
For this reason, the effect of
a relatively large $M_s$, combined with weak equilibration and
the neutrality constraints, is to put a stress on the CFL pairing 
pattern: these effects would all act to pull apart the Fermi momenta
of the different flavors by an amount of order $M_s^2/\mu$
in the absence of CFL pairing.    
(This can be seen by an analysis of neutral unpaired quark matter
in which the Fermi momenta of the $d$, $u$ and $s$
quarks are split by $\simeq M_s^2/4\mu$.)
In the CFL phase, Fermi momenta do not separate~\cite{Rajagopal:2000ff} 
but the consequence
of the stress is that
the excitation energies of those fermionic quasiparticles
whose excitation would serve to ease the stress by
breaking pairs and separating Fermi surfaces
is reduced, again
by of order $M_s^2/\mu$~\cite{Alford:2003fq}. 
When the density becomes
low enough, some of the quasiparticles become gapless,
the CFL pairing pattern is disrupted,
and we enter the gapless CFL (gCFL) 
phase~\cite{Alford:2003fq,Alford:2004hz}, the second-densest
phase on the QCD phase diagram.
Since the strength of the CFL pairing is measured by the gap
parameter $\De_{CFL}$, and the stress
on it is of order $M_s^2/\mu$, the CFL pattern ``breaks'' 
and the CFL$\to$gCFL transition occurs when
the density is low enough that $M_s^2/\mu \sim \Delta_{CFL}$.
Making this argument quantitative results in the prediction
of a $T=0$ second-order insulator-metal transition 
separating the CFL and gCFL phases 
at $M_s^2/\mu \simeq 2 \Delta_{CFL}$~\cite{Alford:2003fq,Alford:2004hz}.
An analogous zero temperature metal insulator transition
has been analyzed in Ref.~\cite{Liu:2004mh}.
(If the CFL phase is augmented
by a $K^0$-condensate~\cite{Bedaque:2001je,Forbes:2004ww}, 
the CFL$\rightarrow$gCFL transition 
is delayed to a value of $M_s^2/\mu$ that is
higher by a factor of $4/3$~\cite{Kryjevski:2004jw}
or less~\cite{Forbes:2004ww}.)

\subsection{The nature of gCFL pairing}

In the gCFL phase the pairing is still antisymmetric in flavor
as well as color and spin, so as in CFL there are diquark condensates
(gap parameters)
$\De_1 \sim \<ds\>$, $\De_2\sim \<us\>$, $\De_3\sim \<ud\>$.
But unlike the CFL phase, where the gap parameters
are very similar, $\De_1=\De_2\simeq\De_3$, in gCFL the
pairing involving strange quarks is suppressed:
very strongly for $\<ds\>$, and quite strongly for $\<us\>$,
so that $\De_1 < \De_2 < \De_3$,
as shown in Figures in 
Refs.~\cite{Alford:2003fq,Alford:2004hz,Fukushima:2004zq}. The 
result is that while
quarks of all three colors and all three flavors still form
Cooper pairs, there are regions of momentum space in
which there is no $\<ds\>$ pairing, and other (very narrow)
regions in which there is no $\<us\>$ pairing,
and these regions are bounded by momenta at which the relevant
fermionic quasiparticles are gapless. 

The gCFL phase is an electromagnetic conductor: unlike the CFL
phase, it contains electrons~\cite{Alford:2003fq,Alford:2004hz}.
The electron chemical potential $\mu_e$ increases as 
$M_s^2/\mu$ is increased, rising from zero at
the CFL$\rightarrow$gCFL transition to values
which are comparable to or even larger than its typical values in
unpaired quark matter, which has $\mu_e\simeq M_s^2/4\mu$.

\subsection{The gCFL domain}

To discuss the range of densities over which gCFL is expected
to occur, we shall parametrize the strength
of the attractive interaction between quarks by 
$\De_0$, which we define as the value of the CFL gap parameter
at $M_s=0$ in quark matter with $\mu=500$~MeV.
(We shall quote all numerical results at $\mu=500~\MeV$,
corresponding to baryon densities between $8.8~n_0$ 
and $9.1~n_0$ depending on the value of $\De_0$ that we choose, 
where $n_0=0.17~{\rm fm}^{-3}$ is the baryon density in nuclear matter.) 
Because asymptotic-density calculations are not quantitatively
valid at this $\mu$, $\De_0$ is not known precisely, with estimates
ranging from 10 to 100 MeV~\cite{Reviews}.

The gCFL phase extends over a range of $M_s^2/\mu$ from
the continuous
CFL$\rightarrow$gCFL transition at $M_s^2/\mu=2\De_1\simeq 2\De_0$
up to a first order phase transition where gCFL gives way to
some phase with even less pairing.  In model calculations, this
transition occurs at $M_s^2/\mu\simeq 5\De_0$, although
this is only quantitatively determined within particular 
models~\cite{Alford:2003fq,Alford:2004hz,Ruster:2004eg,Fukushima:2004zq}.  
To give a sense of the scales involved, for $\De_0=25~\MeV$
and $M_s=250~\MeV$, the gCFL window $2\De_0\lesssim M_s^2/\mu
\lesssim 5\De_0$
corresponds to $320~\MeV\lesssim \mu\lesssim 800~\MeV$. At the lower end of
this range in $\mu$ (upper end in $M_s^2/\mu$) hadronic
matter would be more favorable than any form of quark matter.
And, the upper end of this range in $\mu$ (lower end in $M_s^2/\mu$)
corresponds to densities much higher than those achievable
in neutron stars.  Hence, with these choices of parameters
all the quark matter within neutron stars 
would be in the gCFL phase.
For larger
$\De_0$ or smaller $M_s$, the gCFL window shifts to lower
$\mu$, and neutron stars with a CFL core surrounded by
a gCFL layer become possible.  In reality, both $\De_0$
and $M_s$ are $\mu$-dependent, making these estimates
illustrative only.

\subsection{Gapless quasiparticles in the gCFL phase}

\begin{figure}[t]
\begin{center}
\includegraphics[width=0.48\textwidth]{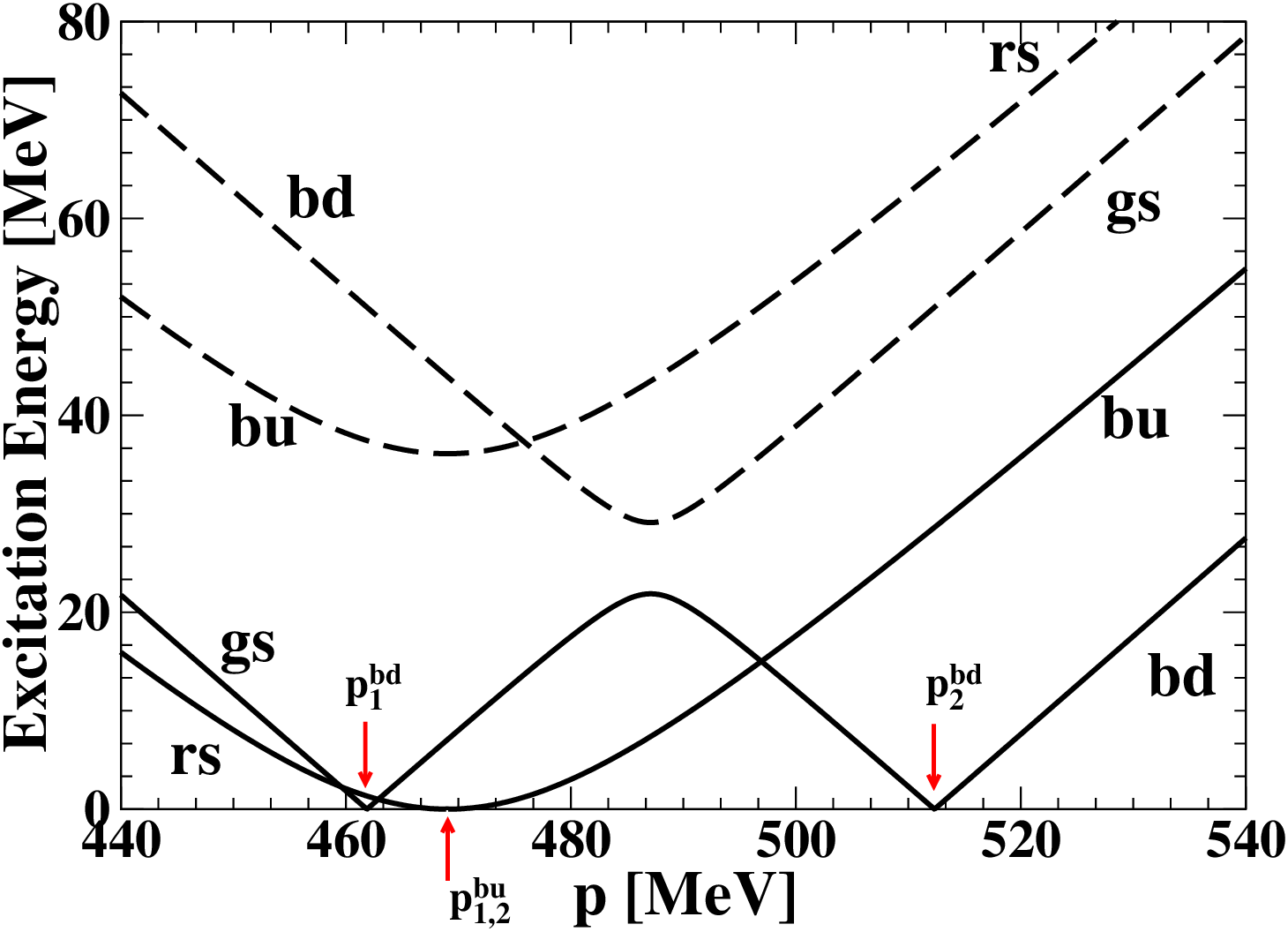}
\end{center}
\vspace{-0.25in}
\caption{Dispersion relations for quasiquarks with
$gs$-$bd$ pairing ($\De_1=3.6~\MeV$) 
and $bu$-$rs$ pairing ($\De_2=18.5~\MeV$),
in the model calculation of 
Refs.~\cite{Alford:2003fq,Alford:2004hz,Fukushima:2004zq} done
at $\mu=500~\MeV$, $M_s^2/\mu=100~\MeV$, with $\De_0=25~\MeV$.
We find gapless $gs$-$bd$ modes at
$p_1^{bd}=461~\MeV$ and $p_2^{bd}=512~\MeV$.
One $bu$-$rs$ mode is gapless
with an almost exactly quadratic dispersion relation.
Actually it is gapless at two momenta
$p_1^{bu}$ and $p_2^{bu}$, but these are too close
together to be resolved until the temperature drops below the eV scale,
meaning we can treat them as
a single zero at $p_{1,2}^{bu}=469~\MeV$.
The five quark 
quasiparticles not plotted are all fully gapped in
the CFL and gCFL phases.
}
\label{fig:disprel}
\end{figure}

In the CFL phase, all nine fermionic quasiparticles are gapped. 
In the gCFL phase, two dispersion relations are 
gapless, as shown in Fig.~\ref{fig:disprel}. We label the
three quark colors as $r$,$g$,$b$, and make the
by now conventional choice for which colors pair with
which flavors in the CFL phase. In this notation
one of the gapless branches describes quasiparticle
excitations that are superpositions
of $bd$ particles and $gs$ holes.
These excitations are gapless
at two momenta $p_1^{bd}$ and $p_2^{bd}$ 
shown in Fig.~\ref{fig:disprel} and
given by~\cite{Alford:2004hz}
\begin{equation}
\half(\mu_{gs}+\mu_{bd})\pm\sqrt{\left[\half\left(\mu_{gs}-\mu_{bd}
\right)\right]^2-\De_1^2}\ ,
\end{equation}
where $\mu_{gs}$ and $\mu_{bd}$ are determined by the 
(nontrivial) requirements of color
and electric neutrality. They are defined in Ref.~\cite{Alford:2004hz},
and their values as a function of $M_s^2/\mu$ 
at various $\De_0$ can be determined from 
plots in Refs.~\cite{Alford:2004hz,Fukushima:2004zq}.
The gapless excitations at $p_1^{bd}$ are predominantly
$gs$, with $bd$ contributing only a small component in
the superposition.  Those at $p_2^{bd}$ are predominantly
$bd$. 
In Section \ref{sec:neutrino_emissivity} 
we shall focus on this dispersion relation
in the vicinity of $p_1^{bd}$, where it 
takes the form
\begin{equation}
\epsilon_{bd}(p)=v_{bd}|p-p_1^{bd}|\ ,
\label{lindisp}
\end{equation}
where 
\begin{equation}
v_{bd}=\sqrt{1-\frac{\Delta_1^2}{\left[\half\left(\mu_{gs}-\mu_{bd}
\right)\right]^2}}
\label{fermivelocity}
\end{equation}
is the Fermi velocity of the gapless quasiparticles.
The $bd$ states with momenta between 
$p_1^{bd}$ and $p_2^{bd}$ are filled, whereas
the $gs$ states in this momentum range are empty.
This means that there is no $gs$-$bd$ pairing in 
the ground state wave function in this region of
momentum space,
although (as the dispersion relations show) there
is still pairing among the excitations. 
The width $p_2^{bd}-p_1^{bd}$ of 
this ``blocking region'' wherein pairing is ``breached''
is zero at the CFL$\rightarrow$gCFL phase transition, and grows
steadily with increasing $M_s^2/\mu$ throughout the gCFL phase.
These dispersion relations behave similarly to those
describing the gapless modes in the two-flavor gapless
2SC phase~\cite{Shovkovy:2003uu}, and in a metastable
three-flavor phase discovered in an early analysis in
which the constraints imposed by neutrality were not
considered~\cite{Alford:1999xc}.

The physics of the gapless $bu$-$rs$ dispersion relation is
interestingly different.  As above these excitations are gapless at
two momenta $p_1^{bu}$ and $p_2^{bu}$ given by
\begin{equation}
\half(\mu_{bu}+\mu_{rs})\pm\sqrt{\left[\half\left(\mu_{bu}-\mu_{rs}
\right)\right]^2-\De_2^2}\ ,
\end{equation}
but in this instance $[\half(\mu_{bu}-\mu_{rs})]^2-\De_2^2$ is {\it very}
small, making the dispersion relation in Fig.~\ref{fig:disprel}
look quadratic with
a single zero at $p_{1,2}^{bu}=\half(\mu_{bu}+\mu_{rs})$. 
For the parameters of Fig.~\ref{fig:disprel},
$p_2^{bu}-p_1^{bu}=0.026~\MeV$ and the height of the dispersion
relation half way between these very nearby gapless points
is only about $5~{\rm eV}$.  Since this is much smaller than the
temperatures that will be of interest to us, we can
safely treat the dispersion relation as quadratic, with
a dispersion relation in the vicinity of its gapless point
given approximately by
\begin{equation}
\epsilon_{bu}(p)= \frac{\left(p-p_{1,2}^{bu}\right)^2}{2\De_2}\ ,
\label{quaddisp}
\end{equation}
with the velocity $v_{bu}$, defined analogously to $v_{bd}$ of
(\ref{fermivelocity}), vanishing at the gapless point. 
The gap parameter $\De_2$ and the
chemical potentials that determine $p_{1,2}^{bu}$
are plotted
as functions of $M_s^2/\mu$ at several values
of $\De_0$ in Refs.~\cite{Alford:2004hz,Fukushima:2004zq}.

This near-quadratic dispersion relation is not a result of fine tuning.
It occurs at all $\mu$ in the gCFL phase, and arises from the fact that
bulk matter must be electrically and color
neutral~\cite{Alford:2003fq,Alford:2004hz}.    In both the CFL and
gCFL phase, there is an unbroken gauge symmetry, 
denoted $U(1)_{\tilde Q}$,  
generated by a linear combination of the generators of
electromagnetic and color symmetry~\cite{Alford:1998mk}. 
Among the neutrality constraints,
it is the imposition of $\tilde Q$-neutrality
that has the implication of interest.  The quarks in the gCFL phase
are almost $\tilde Q$-neutral by themselves, but not quite: their
small excess positive $\tilde Q$ charge is cancelled by
a small admixture of electrons, which have $\tilde Q=-1$.
The excess of 
unpaired $bd$-quarks, occurring in
a broad band of momenta $p_1^{bd}<p<p_2^{bd}$, does not
contribute to the $\tilde Q$ imbalance because these quarks
have $\tilde Q=0$.
It is the unpaired $bu$-quarks with $p_1^{bu}<p<p_2^{bu}$ that 
matter, because they have $\tilde Q=+1$.  They contribute
a positive $\tilde Q$-charge density of order 
$\mu^2(p_2^{bu}-p_1^{bu})$, balanced by the electron
number density, of order $\mu_e^3$. Since $\mu_e\ll \mu$
throughout the gCFL phase,
$p_2^{bu}-p_1^{bu}$ is forced (by the dynamics of the gauge fields
that maintain neutrality) to remain extremely small,
parametrically of order $\mu_e^3/\mu^2$, throughout the gCFL phase.

As described above, the dispersion
relations for the $gs$-$bd$ quasiparticles
are linear about their gapless momenta
$p_1^{bd}$ and $p_2^{bd}$, as in Eq.~(\ref{lindisp}), at generic 
values of $M_s^2/\mu$ in the gCFL phase.  However,
this dispersion relation is fine-tuned to be quadratic
precisely at the CFL$\rightarrow$gCFL transition, where
$p_1^{bd}=p_2^{bd}$ and the Fermi velocity $v_{bd}=0$.
For values of $M_s^2/\mu$ that are in the gCFL regime
but are close to the CFL$\rightarrow$gCFL transition, therefore,
the simplified linear
expression in Eq.~(\ref{lindisp}) cannot be used.
Indeed, if we are interested in those excitations with
energies of order $T$ or less, 
the linear expression
in Eq.~(\ref{lindisp}) is a good approximation as long as
$v_{bd} \gtrsim \sqrt{2T/\De_1}$.
And, close enough to 
the transition that
$v_{bd}\ll \sqrt{2T/\De_1}$
the $gs$-$bd$ dispersion relation can be approximated as
quadratic, as is appropriate for the $bu$-$rs$ dispersion relation
throughout the gCFL phase.

The gapless excitations of the gCFL phase whose dispersion
relations we have described determine the specific heat
and neutrino emissivity of this phase of matter. In the
next two sections, we calculate these quantities in turn.

\section{Specific Heat of Gapless CFL Quark Matter}
\label{sec:specific_heat}
 
The specific heat of any phase of matter
is essentially a count
of the number of possible excitations
with excitation energies of order $T$ or smaller.
Precisely, the contribution of quasiparticle
excitations with dispersion relation $\epsilon(p)$
to the specific heat (heat capacity per unit volume)
at temperature $T$ is given by
\begin{eqnarray}
c_V &=& 2\int \frac{d^3p}{(2\pi)^3} \,\epsilon(p) 
\frac{d}{dT}\left( \frac{1}{e^{\epsilon(p)/T}+1} \right)\nonumber\\
&=&\frac{2}{T^2}\int \frac{d^3p}{(2\pi)^3}\, 
\frac{\epsilon(p)^2}{\left(e^{\epsilon(p)/T}+1\right)
\left(e^{-\epsilon(p)/T}+1\right)}
\label{cVdefn}
\end{eqnarray}
where the prefactor 2 assumes that the quasiparticle
is doubly degenerate by virtue of its spin.  Clearly,
only those excitations with $\epsilon \lesssim T$ 
are important.  For the gapless quasiparticle with 
quadratic dispersion relation, $\epsilon$ is near zero
for $p$ near $p_{1,2}^{bu}$
and in this regime the dispersion relation
can be approximated as in (\ref{quaddisp}).
Approximating the dispersion relation in this way
will give us the specific heat in the
small $T$ limit, and so although it
is straightforward to obtain an ``exact'' result upon
assuming (\ref{quaddisp}), we simply quote the leading
result in the small $T$ limit:
\begin{eqnarray}
c_V&=&\frac{3 (\sqrt{2}-1)\zeta\left(\threehalves\right)}{4\pi^{3/2}}
\, \left(p_{1,2}^{bu}\right)^2 \De_2^{1/2} T^{1/2}\nonumber\\
&\simeq& 0.146 \left(p_{1,2}^{bu}\right)^2 \De_2^{1/2} T^{1/2}\ .
\label{quadcV}
\end{eqnarray}
As expected, 
this is proportional to 
the number of excitations with energy less than $T$, given
a quadratic dispersion relation (\ref{quaddisp}). 
For quasiparticles with conventional linear dispersion relations,
namely $\epsilon(p)=v|p-p_F|$ for some $p_F$ and $v$, the
expression (\ref{cVdefn}) yields the familiar $c_V=\third p_F^2 T/v$.
Hence, 
the specific heat of the gCFL phase is given by
\begin{widetext}
\beq
c_V = \frac{k_B}{(\hbar c)^3}\left[
0.146 \left(p_{1,2}^{bu} c \right)^2 \De_2^{1/2} (k_B T)^{1/2}
+ \frac{c}{3v_{bd}} \left(p_1^{bd} c \right)^2 k_B T 
+\frac{c}{3v_{bd}} \left(p_2^{bd} c \right)^2 
k_B T 
+\third \mu_e^2 k_B T \right] \ ,
\label{gCFLspecificheat}
\eeq
\end{widetext}
where we have restored factors of $\hbar$, $c$ and Boltzmann's constant $k_B$
and 
where the last term comes from the electrons and is 
negligible because   all the quark Fermi momenta are of order $\mu$,
and $\mu_e\ll\mu$. 
As long as $T\ll \De_2$, the contribution from the
quasiparticle with quadratic dispersion relation dominates.
We shall describe reasonable values of $\De_2$ and
$T$ in subsequent sections; it 
suffices here to say that $\De_2/T$ is of order
hundreds or thousands.

Note that close enough to the CFL$\rightarrow$gCFL
transition that $v_{bd}\lesssim \sqrt{2T/\De_1}$, the 
$gs$-$bd$ dispersion relation cannnot be treated as linear,
and the expression (\ref{gCFLspecificheat}) is modified.
Indeed, if $v_{bd}\ll \sqrt{2T/\De_1}$ the $gs$-$bd$ dispersion
relation can be treated as quadratic, making their contribution
to the specific heat comparable to that of the $bu$-$rs$
quasiparticles.

The gCFL phase also has light bosonic excitations, 
for example that associated with superfluidity, but their
contribution to the specific heat is of order $T^3$ and
so can be neglected.   The specific heat may be
enhanced by logarithimc corrections analogous to those in unpaired
quark matter~\cite{Ipp:2003cj}, but we leave
their analysis to future work.

\section{Neutrino Emissivity of Gapless CFL Quark Matter}
\label{sec:neutrino_emissivity}

The presence of gapless quark quasiparticles 
in the gCFL phase raises
the possibility of neutrino emission by direct Urca
processes.  Because the gapless modes are superpositions
of $gs$ and $bd$ quarks, and $bu$ and $rs$ quarks,
and because weak interactions cannot change the color
of a quark, the direct Urca processes that we must
consider are
\beq
bd \rightarrow bu + e^{-} + \bar{\nu}
\label{process1}
\eeq
and
\beq
bu + e^{-} \rightarrow bd + \nu\ .
\label{process2}
\eeq
Momentum conservation in these reactions requires
that the momenta of the two quarks and the electron
form a triangle~\cite{Iwamoto}. (The argument is that these three
fermions must all have energies of order $T$ and
hence must all be close to the momenta at which
their dispersion relations are gapless. The
neutrinos escape from the star, and hence
have zero Fermi momentum. By energy
conservation, the escaping neutrino therefore has energy,
and hence momentum, of order $T$. This is negligible
compared with the momenta of the other fermions.)
Momenta of the two quarks and the electron satisfying
the triangle constraint can only be found if 
$|p^{bu}-p^{bd}|\leq \mu_e$, where $p^{bu}$
and $p^{bd}$ are the magnitudes of the momenta at which the 
$bu$ and $bd$ quarks are gapless.  The only momentum
at which gapless quasiparticles with a $bu$ component
are found is $p_{1,2}^{bu}$.
Gapless $bd$
quarks occur at two momenta.
For the parameters
in Fig.~\ref{fig:disprel}, the electron Fermi momentum
is $\mu_e=25.9~\MeV$, meaning that the triangle constraint
can be satisfied if we choose $bd$ quarks with momenta
near $p_1^{bd}$, but cannot be satisfied for those near
$p_2^{bd}$.  Indeed, we find that direct Urca processes involving the
quasiparticles at $p_2^{bd}$ are forbidden
throughout the gCFL regime of $M_s^2/\mu$,
whereas those involving quasiparticles at $p_1^{bd}$
are allowed throughout all of the gCFL regime, with
the available phase space vanishing at the CFL$\rightarrow$gCFL
transition and opening up with increasing $M_s^2/\mu$.

The calculation of the neutrino emissivity due to direct Urca
processes in unpaired quark matter was first 
done by Iwamoto in Ref.~\cite{Iwamoto}, and we shall
follow his analysis,
leaving the calculation of any logarithmic enhancement
analogous to that in unpaired quark matter~\cite{Schafer:2004jp}
to future work.  
There are two essential differences between Iwamoto's
calculation for unpaired quark matter and ours for the gCFL phase.
First, although the quasiparticles near $p_1^{bd}$ have
a conventional linear dispersion relation $\epsilon_{bd}(p)$
given in (\ref{lindisp}), as in unpaired quark matter, 
the quasiparticles near $p_{1,2}^{bu}$ have a quadratic
dispersion relation $\epsilon_{bu}(p)$ given in (\ref{quaddisp}).
Analogous to its effect on the specific heat, this
unusual dispersion relation
increases the available phase space for the direct Urca reactions
by a factor of order $\sqrt{\De_2/T}$ relative to 
that in the standard
calculation, resulting in a neutrino emissivity
$\varepsilon_\nu\sim T^{5.5}$ rather than $\sim T^6$.
Second, the quasiparticles near 
$p_1^{bd}$ and $p_{1,2}^{bu}$ with dispersion
relations $\epsilon_{bd}$ and $\epsilon_{bu}$ 
are {\it not} purely $bd$ and $bu$ quarks.
They are superpositions of $bd$ and $gs$ quarks, and $bu$ and $rs$
quarks, respectively.  
Only the  $bd$ and $bu$ components of the quasiparticles
participate, meaning that the neutrino emissivity
is proportional to the probabilities that each quasiparticle
is blue. These probabilities are the
squares of the Bogoliubov coefficients $B_{bu}$
and $B_{bd}$, specified as follows~\cite{Alford:1998mk}. 
$B_{bu}$ is given by
\beq
B_{bu}(p)^2=\frac{1}{2}\left(1+\frac{p-p_{1,2}^{bu}}
{\sqrt{(p-p_{1,2}^{bu})^2+ \Delta_2^2}}\right)
\eeq
where $p_{1,2}^{bu}=\half(\mu_{bu}+\mu_{rs})$ and
where $\Delta_2$ is the gap parameter for the pairing 
between $bu$ and $rs$ quarks.
To simplify the calculation we Taylor expand the coefficient 
around $p_{1,2}^{bu}$ and get
\beq
B_{bu}(p)^2=\frac{1}{2}\left(1+\frac{p-p_{1,2}^{bu}}{\Delta_2}\right)\ . 
\label{firstbogo}
\eeq
The $bd$ Bogoliubov coefficient is given by
\beq
B_{bd}(p)^2=\frac{1}{2}\left(1+ \frac{p-\bar{\mu}}{\sqrt{(p-\bar{\mu})^2 
+ \Delta_1^2}}\right)
\eeq
where $\bar\mu=\half(\mu_{bd}+\mu_{gs})$ and
where $\Delta_1$ is the gap parameter for the pairing 
between $bd$ and $gs$ quarks.
Because
the dispersion relation is linear, the quarks that contribute to the
emissivity lie in a band about $p_1^{bd}$
whose width is only of order $T$,
and we shall replace $B_{bd}(p)$ by $B_{bd}\left(p_1^{bd}\right)$.
This coefficient can be quite small. For example, with
parameters as in Fig.~\ref{fig:disprel},
meaning in particular $M_s^2/\mu=100~\MeV$, 
the probability that the gapless quasiparticles
at $p_1^{bd}$ are in fact $bd$ is only 
$B_{bd}\left(p_1^{bd}\right)^2=0.00479$.

We now present the calculation of $\varepsilon_{\nu}$, following
Ref.~\cite{Iwamoto}.
The transition rate for the process (\ref{process1}) is
\beq
W=\frac{V(2\pi)^4 \delta^{(4)}(p_{bd} - p_{\nu} - p_{bu} - p_e)}
{\prod_{i=1}^4 2 E_iV}
\left| M \right|^2 
\label{w}
\eeq
where the 
index $i$ runs over the four species that participate in the interaction. 
V is the normalization volume, which will drop out
by the end of the calculation,  and
the squared amplitude $ \left| M \right|^2$ is given by
\beq
\left|M \right|^2
= 64G_F^2 \cos^2 \theta_c (p_{bd} \cdot p_{\nu})(p_{bu} \cdot p_e)\ ,
\label{m}
\eeq
where we have averaged over 
the spin of the initial down quark and summed 
over the spin of the final up quark.
Here, $G_F$ is the Fermi constant and $\theta_c$ is the Cabibbo angle.

The neutrino emissivity is the rate of energy loss per unit volume
due to neutrino emission. 
It is obtained by multiplying the transition rate by
the neutrino energy and integrating over the available
phase space, weighted by the Bogoliubov coefficients. The
expression can be written as
\begin{widetext}
\beq
\varepsilon_\nu = \frac{2}{V} \left[ 
\prod_{i=1}^4 V \int \frac{d^3p_i}{(2\pi)^3}\right] 
E_{\nu}\,W\,n(p_{bd})(1-n(p_{bu}))(1-n(p_e)) 
B_{bu}(p_{bu})^2 B_{bd}(p_1^{bd})^2\ .
\label{emis}
\eeq
\end{widetext}
Here, the 
Fermi distribution functions $n(p_{bd})(1-n(p_{bu}))(1-n(p_e))$ 
state that in order for the process to occur
we have to have an occupied down 
state and unoccupied up and electron states. 
In thermal equilibrium (which is maintained by strong
and electromagnetic processes occurring on timescales much
faster than neutrino emission via weak interactions)
the distribution functions are given by
\beq
n(p_i)=\frac{1}{1+\exp x_i}
\label{distrib}
\eeq
where
\beq
x_e=\frac{p_e - \mu_e}{T}\ ,
\eeq
where
\beq
x_{bd}
=\pm \frac{\epsilon_{bd}(p_{bd})}{T}\ ,
\eeq
with the $\pm$ serving in effect to undo the absolute
value in (\ref{lindisp}),
and where
\beq
x_{bu}=\pm \frac{\epsilon_{bu}(p_{bu})}{T}
\eeq
with the $\pm$ chosen positive
for $p_{bu}>p_{1,2}^{bu}$ and negative for  $p_{bu}<p_{1,2}^{bu}$. 
In defining $x_{bd}$ we have used (\ref{lindisp}), meaning
that this derivation is valid at generic values of $M_s^2/\mu$
in the gCFL regime where $v_{bd}\gtrsim \sqrt{2T/\De_1}$, but not 
close to the CFL$\rightarrow$gCFL transition, where 
both the $bd$ and $bu$ branches should be treated
as quadratic.  We shall discuss
this further below.
For later use, we also define
\beq
x_\nu = \frac{p_\nu}{T}\ .
\eeq
It is possible to set the calculation up directly in terms
of the positive quasiparticle excitation energies,
but introducing the $\pm$ as we have done allows us
to follow Iwamoto's calculation more closely.

We combine Eqs. (\ref{w}), (\ref{m}), (\ref{emis}), (\ref{distrib}) 
and multiply
by a factor of 2 in order to include the emissivity due
to the second process (\ref{process2}), whose
contribution proves to be the same as that above. 
We write the integration element $d^3p_i=p_i^2 dp_i d \Omega_i$, where 
$d\Omega_i$ is the infinitesimal solid angle. The complete 
expression for the emissivity then takes the form~\cite{Iwamoto}
\beq
\varepsilon_\nu = \frac{G_F^2}{16\pi^8}\cos^2 \theta_c (1-\cos \theta_{ue})
{\cal A}{\cal B}\ .
\label{ABdefinition}
\eeq
Here, ${\cal A}$ is an angular integral defined as
\beq
{\cal A}=\left(\prod_{i=1}^{4} \int d\Omega_i\right)
\delta({\bf p}_{bd}-{\bf p}_{bu}-{\bf p}_e)\ ,
\eeq
where we have eliminated certain terms that vanish identically
upon angular integration.  ${\cal A}$ 
is identical to that in Ref.~\cite{Iwamoto} and upon
evaluation yields
\beq
{\cal A}=\frac{32 \pi^3}{p_1^{bd}p_{1,2}^{bu}\mu_e}\ ,
\label{Aresult}
\eeq
where we have taken $|{\bf p}_{bd}|=p_1^{bd}$, 
$|{\bf p}_{bu}|=p_{1,2}^{bu}$, $|{\bf p}_{e}|=\mu_e$, knowing
that these are the values at which the ${\cal B}$ integral
is dominated. The integral ${\cal B}$ is defined as
\begin{widetext}
\begin{eqnarray}
{\cal B}&=&\int_{0}^{\infty}p_{bd}^2dp_{bd} 
\int_{0}^{\infty}p_{bu}^2dp_{bu} 
\int_{0}^{\infty}p_e^2dp_e 
\int_{0}^{\infty}p_{\nu}^3dp_\nu
n(p_{bd})(1-n(p_{bu}))(1-n(p_e))\nonumber\\
&~& \qquad\qquad\qquad\qquad\qquad\qquad\qquad\qquad\qquad\qquad
\times\frac{1}{T}\delta\left(x_{bd}-x_{\nu}- x_{bu}-x_e\right)
B_{bu}(p_{bu})^2 B_{bd}(p_1^{bd})^2\ .
\end{eqnarray}
Finally, the angle $\theta_{ue}$ in (\ref{ABdefinition}) 
is the angle between the momentum
of the $bu$-quark and that of the electron, when the two quark
momenta and the electron momentum are arranged in a momentum conserving
triangle. A little trigonometry shows that 
$\theta_{ue}=\theta_{de}+\theta_{du}$ where
\beq
\cos \theta_{de}=\frac{(p_1^{bd})^2 + \mu_e^2 -(p_{1,2}^{bu})^2 }
{2p_1^{bd}\mu_e}
\eeq
and
\beq
\cos \theta_{du}=\frac{ (p_1^{bd})^2-\mu_e^2+(p_{1,2}^{bu})^2 }
{2p_1^{bd}p_{1,2}^{bu}}\ .
\eeq
Now, all that remains is the evaluation of ${\cal B}$.

The integral ${\cal B}$ is dominated by $p_{bd}$ near $p_1^{bd}$,
by $p_{bu}$ near $p_{1,2}^{bu}$, and by $p_e$ near $\mu_e$
so we can pull the factor $p_{bd}^2 p_{bu}^2 p_e^2$ out of the
integrand and replace  it by $(p_1^{bd})^2 (p_{1,2}^{bu})^2 \mu_e^2$.
Next, we change variables of integration from the $p$'s to
the $x$'s and obtain 
\begin{eqnarray}
{\cal B}&=&(p_1^{bd})^2 (p_{1,2}^{bu})^2 \mu_e^2 T^6 
\frac{B_{bd}(p_1^{bd})^2}{2} 
\int_{-\infty}^{\infty}\frac{dx_{bd}}{v_{bd}}\int_{-\infty}^{\infty}dx_e
\int_{0}^{\infty}dx_{\nu}x_{\nu}^3
\int_{-\infty}^{\infty} dx_{bu}
\frac{\sqrt{\Delta_2}}
{\sqrt{2 T \left| x_{bu}\right|}} \frac{1}{e^{x_{bd}}+1} \frac{1}{e^{x_{bu}}+1} 
\frac{1}{e^{x_e}+1}\nonumber\\
&~&\qquad\qquad\qquad\qquad\qquad
\qquad\qquad\qquad\qquad\qquad\qquad
\qquad\qquad\qquad
\times\delta(x_{bd}+x_{bu}+x_e-x_{\nu})
\label{Bsecond}
\end{eqnarray}
where, as in the calculation
of the specific heat, the enhanced density of states 
$dp_{bu}/dx_{bu}=\left(\De_2 T/2\left|x_{bu}\right|\right)^{1/2}$
for the quasiparticle with the quadratic dispersion relation
is crucial.  In (\ref{Bsecond}) we have
made the approximation $B_{bu}(p_{1,2}^{bu})^2\simeq\half$, since
the other  term in (\ref{firstbogo}) leads to a contribution
proportional to $T^6$ and we are keeping only the leading
contribution, proportional to $T^{5.5}$.  We now rewrite (\ref{Bsecond})
as 
\begin{eqnarray}
{\cal B}
&=&\frac{1}{2\sqrt{2}}(p_1^{bd})^2 (p_{1,2}^{bu})^2 \mu_e^2
\frac{T^{5.5}\sqrt{\Delta_2 } B_{bd}(p_1^{bd})^2}{v_{bd}} \int_{-\infty}^{\infty}dx_{bd}
\int_{-\infty}^{\infty}dx_e\int_{0}^{\infty}dx_{\nu}x_{\nu}^3
\int_{-\infty}^{\infty}dx_{bu}\frac{1}{\sqrt{\mid x_{bu}\mid}} 
\frac{1}{e^{x_{bd}}+1} \frac{1}{e^{x_{bu}}+1} \frac{1}{e^{x_e}+1}\nonumber\\ 
&~&\qquad\qquad\qquad\qquad\qquad
\qquad\qquad\qquad\qquad\qquad\qquad
\qquad\qquad\qquad\qquad\qquad\qquad
\times\delta(x_{bd}+x_{bu}+x_e-x_{\nu})\ ,
\end{eqnarray}
use the delta function to perform one of 
the integrations, and then 
perform the remaining dimensionless triple integral numerically.
The result is 
\beq
{\cal B}=31.18\, (p_1^{bd})^2 (p_{1,2}^{bu})^2 \mu_e^2
\frac{T^{5.5}\sqrt{\Delta_2 } B_{bd}(p_1^{bd})^2}{v_{bd}} \ .
\eeq
Combining this with the result (\ref{Aresult}) for ${\cal A}$
and substituting into (\ref{ABdefinition})
we obtain the final result for the neutrino emissivity of
the gCFL phase 
\beq
\varepsilon_\nu=\frac{62.36}{\pi^5}\frac{G_F^2 \cos^2 \theta_c}
{\hbar^{10}  c^7}
(1-\cos \theta_{ue} )p_1^{bd} p_{1,2}^{bu} \mu_e \frac{(k_B T)^{5.5}
\sqrt{\Delta_2 } 
B_{bd}(p_1^{bd})^2}{v_{bd}}\ ,
\label{emissivityresult}
\eeq
where we have restored the factors of $\hbar$, $c$ and $k_B$.
This result is valid as long as $v_{bd}\gtrsim \sqrt{2T/\De_1}$, meaning
that the $gs$-$bd$ dispersion relation can be treated as linear
with slope $v_{bd}$.  At any nonzero temperature, there is
a region just
on the gCFL side of 
the CFL $\rightarrow$ gCFL transition where this approximation
breaks down.  Indeed, in the region so close to the transition
that $v_{bd}\ll \sqrt{2T/\De_1}$, the $gs$-$bd$ and $bu$-$rs$ dispersion
relations can both be treated as quadratic, and an analysis similar
to that we have presented above yields
\beq
\varepsilon_\nu=\frac{42.70}{\pi^5}\frac{G_F^2 \cos^2 \theta_c}
{\hbar^{10}  c^7}
(1-\cos \theta_{ue} )p_1^{bd} p_{1,2}^{bu} \mu_e (k_B T)^{5}
\sqrt{\Delta_2 \Delta_1} 
B_{bd}(p_1^{bd})^2\ .
\label{secondemissivityresult}
\eeq
\end{widetext}
We shall see below that for temperatures of interest in neutron
star physics, this expression is valid only in a very narrow 
window of parameter space. It is (\ref{emissivityresult}) that is
relevant to neutron star phenomenology.

The gCFL emissivity (\ref{emissivityresult})
can be compared to the neutrino emissivity of noninteracting
quark matter~\cite{Iwamoto}
\beq
\ba{rcl}
\varepsilon_\nu^{\rm unpaired} &=& \dsp \frac{457 \pi}{1680}
\frac{G_F^2 \cos^2 \theta_c}{\hbar^{10}  c^4}M_s^2 p_F (k_B T)^6 \\[2ex]
&=& ( 3.6 \times 10^{14} ~{\rm erg}\,{\rm cm}^{-3}\,{\rm s}^{-1} ) \times \\
&& \dsp\left( \frac{ (M_s^2/\mu)}{ 100~\MeV} \right)
\left( \frac{\mu}{500~\MeV}\right)^2
\left( \frac{T}{10^7~{\rm K}} \right)^6 \\
\ea
\label{unpairedemissivity}
\eeq
where $p_F\simeq \mu$ is the up quark Fermi momentum and where $\mu_e$
has been replaced by $M_s^2/4\mu$, appropriate for neutral
unpaired quark matter.
The gCFL emissivity (\ref{emissivityresult}) is enhanced by a factor
of $\sqrt{\De_2/T}$ relative to that of noninteracting quark matter,
but the full comparison between the two rates is more involved.

\begin{figure}[t]
\begin{center}
\includegraphics[width=0.49\textwidth]{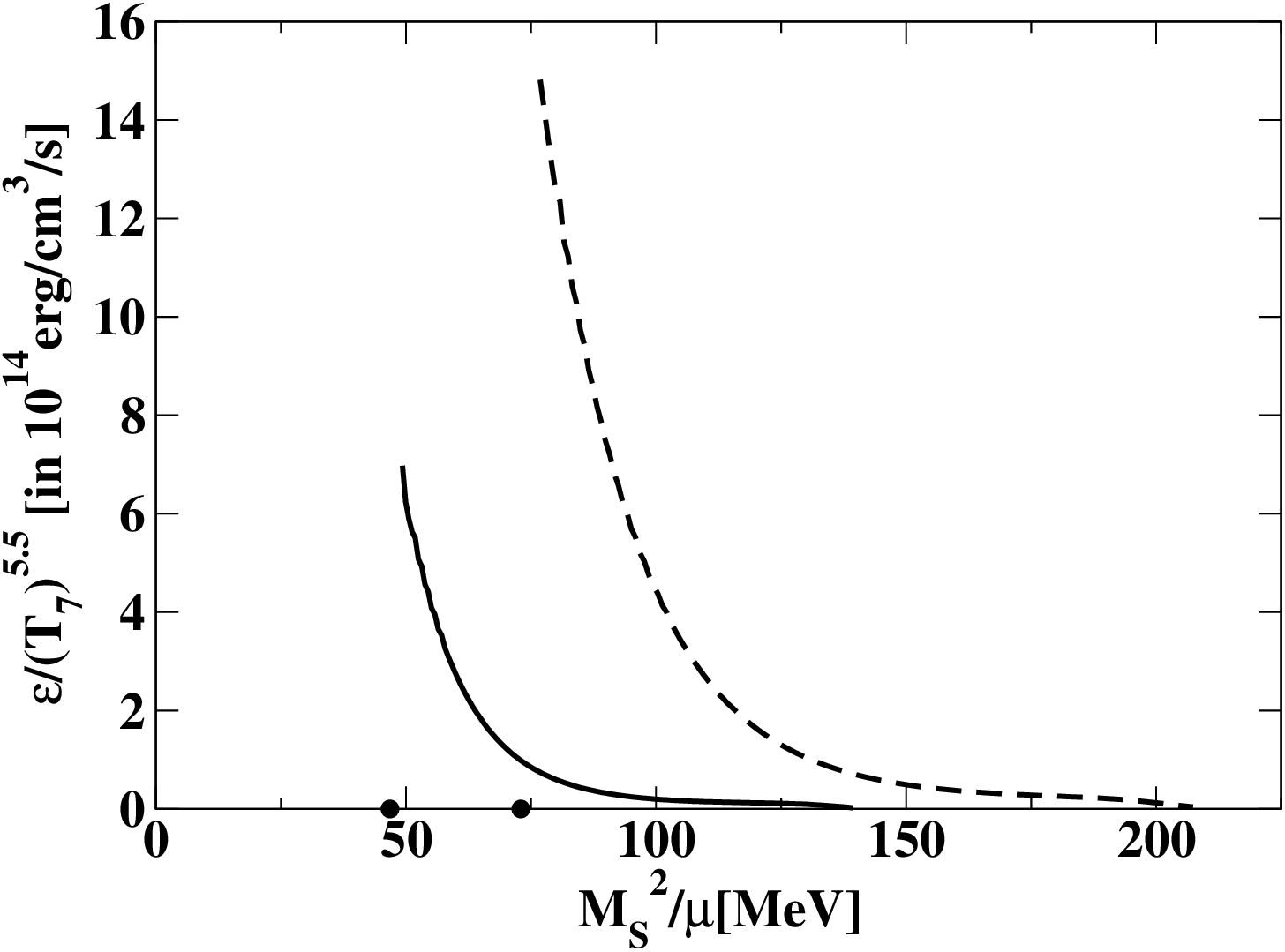}
\end{center}
\vspace{-0.25in}
\caption{Neutrino emissivity $\varepsilon_\nu$ of 
gCFL quark matter from (\ref{emissivityresult}) 
divided by $(T/10^7{\rm K})^{5.5}$,
plotted versus $M_s^2/\mu$.  
The two curves
are drawn for two different values of the strength of
the interaction between quarks, corresponding to 
CFL gap parameters $\De_0=25$ (solid curve) and 
$\De_0=40~\MeV$ (dashed).  The 
location of the CFL $\rightarrow$ gCFL
transitions for $\De_0=25$ and $40~\MeV$ are indicated
by dots on the horizontal axis, at 
$M_s^2/\mu=46.8$ and $73.0~\MeV$ respectively. 
To the left of these dots,
$\varepsilon_\nu$ is negligible in the CFL phase. As discussed
in the text we begin the curves a small interval to the right of the
transition, at the $M_s^2/\mu$ where $v_{bd}=0.15$.
(The $\De_0=25~\MeV$  and $\De_0=40~\MeV$
curves begin $2.5~\MeV$  and $3.9~\MeV$ to the right of their
respective transitions.)
The baryon chemical potential is 
$\mu=500~\MeV$, corresponding to a density about
nine times that of ordinary nuclear matter.  
The effect of changing $\mu$ while keeping $M_s^2/\mu$ 
and $\De_0$ fixed 
can be approximated by scaling $\varepsilon_\nu$ with $\mu^2$.
}
\label{fig:plot2}
\end{figure}

We have obtained our result (\ref{emissivityresult})
in a form which makes the dependence of $\varepsilon_\nu$
on $T$ manifest, but which obscures the
dependence on $M_s$ and $\mu$ because $\De_2$, $B_{bd}(p_1^{bd})$,
$\mu_e$, $v_{bd}$, $p_1^{bd}$ and
$p_{1,2}^{bu}$  all change with $M_s$ and $\mu$.  
The most important $\mu$ dependence is straightforward: 
$p_1^{bd}\sim p_{1,2}^{bu}\sim\mu$ and
hence $\varepsilon_\nu\sim\mu^2$.  
The remaining 
dependence on $M_s$ and $\mu$ is dominated by the dependence
on $M_s^2/\mu$,  
which is nontrivial because $\De_2$,
$B_{bd}(p_1^{bd})$,
$\mu_e$ and $v_{bd}$ and $p_2^{bd}-p_1^{bd}$ 
all depend nontrivially on $M_s^2/\mu$. 
The result also depends on $\De_0$, through this
same set of quantities.  
The reader
who wishes to obtain
numerical values of $\varepsilon_\nu$ would need numerical
values for
the gap parameters and chemical potentials in the gCFL
phase given in plots in 
Refs.~\cite{Alford:2004hz,Fukushima:2004zq}, which are used
in the specification of many of the quantities 
occurring in (\ref{emissivityresult}).  Given all these
implicit dependences, we 
provide Fig.~\ref{fig:plot2}
for the convenience of
the reader who wishes to use our result (\ref{emissivityresult})
for the 
gCFL neutrino emissivity, for example in order to 
calculate its effects on neutron star cooling.

The most important
dependences of $\varepsilon_\nu$ are 
straightforward: $\varepsilon_\nu\sim T^{5.5}$
and $\varepsilon_\nu\sim\mu^2$.  All the remaining
dependences are best described as
dependence on $M_s^2/\mu$ and $\De_0$,
and hence are described by Figure~\ref{fig:plot2}
which shows $\varepsilon_\nu/T^{5.5}$ as a function
of $M_s^2/\mu$ for two values of $\De_0$.  
For each $\De_0$,
we see nonzero neutrino emissivity in the 
corresponding gCFL regime, with $\varepsilon_\nu$ negligible at
lower $M_s^2/\mu$ in the CFL phase.  We do not plot
$\varepsilon_\nu$ at values of $M_s^2/\mu$ that are larger
than the gCFL regime, because it is not known what phase
of quark matter would be favored there, with 
what $T$-dependence for its $\varepsilon_\nu$. 
(Crystalline color superconducting quark matter~\cite{Alford:2000ze}
is a leading candidate for the third-densest phase on
the QCD phase diagram, and its neutrino
emissivity has not been calculated.)  Also, at these
low densities  
quark matter may well have already been superseded by nuclear
matter. 
Note that
a reasonable estimate of the range of $M_s^2/\mu$ of interest
to describe possible quark matter
in neutron stars is $50~\MeV < M_s^2/\mu < 250~\MeV$, corresponding
roughly to $350~\MeV<\mu<500~\MeV$ and $150~\MeV<M_s<300~\MeV$.
If $\De_0$ is large, say $\De_0=100~\MeV$, the curve on 
Fig.~\ref{fig:plot2} shifts far to the right, and
any quark matter that
occurs is likely CFL, with negligible $\varepsilon_\nu$.  We see from
the figure that for $\De_0=25~\MeV$, the highest density
quark matter that can be reached 
is likely in the gCFL phase.  For intermediate values of $\De_0$,
it is possible to obtain a CFL core surrounded by a gCFL layer.

The shape of the curves in Fig.~\ref{fig:plot2} arises
from the combination of many effects.  As 
$M_s^2/\mu$ increases  through the gCFL
regime, $\mu_e$ rises monotonically and  $(1-\cos\theta_{ue})$
initially rises rapidly as phase space for
neutrino emission opens up, and then varies little.  These effects
are overwhelmed by the fact that as $M_s^2/\mu$ increases through the gCFL
regime, $\De_2$, $1/v_{bd}$, and the Bogoliubov coefficient
$B_{bd}(p_1^{bd})$ all decrease monotonically.
%
%
%
%

Close to the CFL $\rightarrow$ gCFL transition, the most important
contribution to the steep decrease of $\varepsilon_\nu$ with
increasing $M_s^2/\mu$ seen in Fig.~\ref{fig:plot2} is
the factor $1/v_{bd}$ occurring in (\ref{emissivityresult}), 
since after all $v_{bd} = 0$
at the transition.   However, one must recall that the
expression (\ref{emissivityresult}) plotted in Fig.~2
is only valid for $v_{bd}\gtrsim \sqrt{2T/\De_1}$.
Indeed, for any nonzero $T$ 
there is a region close to the transition where $v_{bd}\ll \sqrt{2T/\De_1}$
and the emissivity
is given by (\ref{secondemissivityresult}) with $\varepsilon_\nu \sim T^5$,
not by (\ref{emissivityresult}) with $\varepsilon_\nu \sim T^{5.5}$ 
as plotted in Fig.~\ref{fig:plot2}.  
In Fig.~\ref{fig:plot2}, we have 
begun the gCFL curves at the value of $M_s^2/\mu$ 
at which $v_{bd}=0.15$, meaning that
the curves can be trusted as long 
as $T \lesssim \De_1/100 \simeq \De_0/100$. (Note that
$\De_1\simeq\De_0$ near the CFL $\rightarrow$ gCFL transition.)
For typical neutron star temperatures of order keV, 
$\varepsilon_\nu \sim T^{5.5}$ as given by (\ref{emissivityresult}) 
(and the curves of Fig.~\ref{fig:plot2} are therefore valid) 
even closer to
the transition than where we stopped the curves in the Fig~2.
The curves can safely be used in neutron star cooling calculations,
as we shall do in Section~\ref{sec:implications}. 

Further to the right in Fig.~\ref{fig:plot2}, well away from the transition,
$v_{bd}$ approaches 1 and the factor $1/v_{bd}$ ceases to control
the shape of the curves. In this regime,
the most important contribution
to the decline in $\varepsilon_\nu$ is the rapidly falling
Bogoliubov coefficient:  as $M_s^2/\mu$ increases $p_1^{bd}$
and $p_2^{bd}$ in Fig.~\ref{fig:disprel} separate and 
the $bd$-component of the gapless quasiparticle at $p_1^{bd}$,
namely $B_{bd}(p_1^{bd})$, drops faster than
$\mu_e$ rises.

In Section \ref{sec:implications} we shall sketch the implications of our
results for the specific heat and neutrino emissivity
of gCFL quark matter for neutron star cooling. 
The most important dependence of $\varepsilon_\nu$
in this context is its $T$-dependence.
In all plots in Section \ref{sec:implications}, we show two curves,
both with  $M_s^2/\mu=100~\MeV$, one with $\De_0=25~\MeV$
and one with $\De_0=40~\MeV$.  We choose these values
because we see from Fig.~\ref{fig:plot2} that they
correspond to a small, but reasonable, and a larger, but
still reasonable, value of $\varepsilon_\nu/T^{5.5}$.  Were we to
choose values of parameters that happened to
land very close to the CFL $\rightarrow$ gCFL
transition, all the conclusions that we draw in the next section
would become stronger.

\section{Implications for the Cooling of Neutron Stars}
\label{sec:implications}

The central results of this paper are the specific heat
and neutrino emissivity, calculated in Sections \ref{sec:specific_heat}
 and \ref{sec:neutrino_emissivity}.
We shall not attempt a state-of-the-art neutron star cooling
calculation here, preferring instead to provide a calculation
that is better thought of as illustrative, not quantitative. 
The effect that we wish to highlight is both large and qualitative, 
arising from the $T$-dependence of $c_V$ and $\varepsilon_\nu$,
and we expect that even our crude treatment will persuade
the reader of its significance.  

We analyze the cooling
of a ``toy star'' consisting of a volume of ``nuclear matter''
at a constant density $1.5 n_0$ and a volume of denser
quark matter with a constant density specified by 
$\mu=500~\MeV$.  As nuclear matter we take an 
electrically neutral 
gas of noninteracting 
neutrons, protons and electrons in weak equilibrium.
We investigate three different possibilities  for the
quark matter, all 
electrically and color neutral and in weak equilibrium,
and all with $\mu=500~\MeV$ and $M_s^2/\mu=100~\MeV$.
We consider two possibilities for  
quark matter in the gCFL phase, with $\De_0=40~\MeV$
and $\De_0=25~\MeV$. And, we consider noninteracting
quark matter.  These three options have densities
of $9.1$, $8.9$ and $8.8$ times normal nuclear matter  density
$n_0$, respectively. 
By treating the quark matter core as having a constant density, our
calculation neglects the possibility of a
thin spherical 
gCFL-CFL interface region, in which there would be an
enhancement in both the specific heat (by a factor of two, see end of
Sect.~\ref{sec:specific_heat}) 
and the neutrino emissivity 
(see Eq.~(\ref{secondemissivityresult})) relative to the gCFL
expressions (\ref{gCFLspecificheat}) and (\ref{emissivityresult})
that we shall use. Such a shell would be very thin because
these enhancements occur only within a very narrow window 
in $M_s^2/\mu$, but a quantitative investigation of how small its
effects are is not possible in our ``toy star calculation''.
We choose the quark matter and nuclear matter volumes
$V_{qm}$ and $V_{nm}$
such that the total mass of the star is 1.4 solar masses.
If we set the quark matter volume to zero, this corresponds
to choosing a nuclear matter ``star'' that is a sphere
with radius $R=12$~km.  If we include a dense quark matter
core with radius $R_{\rm core}$ while keeping the total
mass fixed, the star shrinks as we
increase $R_{\rm core}$. With $R_{\rm core}=5$~km
the stellar radius is $R=10$~km.   A gCFL core with radius
5 km has the same volume as a gCFL layer extending from 
$r=4.5$~km to $r=6$~km. Since such a layer would surround a
CFL quark matter core, and since CFL matter
plays no role in neutron star
cooling, the estimates that we quote for
$R_{\rm core}=5$~km can equally well be taken as a guide
to this scenario.  
Our final toy-model assumption is
that our ``star'' is a black body.  The work that
needs to be done to turn our illustrative ``toy star
calculation'' into a quantitative calculation of
neutron star cooling includes the investigation of
realistic density profiles, realistic nuclear matter,
and realistic atmospheres.  We defer this, as our calculation
suffices to make our qualitative point.

Our ``star'' loses heat by neutrino emission from its entire
volume and by black body emission of photons from its surface.
The heat loss due to neutrino emission is
\begin{equation}
L_\nu=V_{nm}\varepsilon_\nu^{nm} + V_{qm}\varepsilon_\nu^{qm}\ .
\end{equation}
The quark matter neutrino emissivity $\varepsilon_\nu^{qm}$
is given either
by (\ref{emissivityresult}) or (\ref{unpairedemissivity}), depending on 
whether we are considering a gCFL core or an unpaired
quark matter core.  The nuclear matter emits neutrinos
via modified Urca processes like $n+X\rightarrow p+X+e+\bar\nu$,
with $X$ either a neutron or a proton that serves to carry
away some recoil momentum, in order that momentum and energy
can both be conserved in the process.  The 
resulting emissivity is~\cite{BahcallWolf,ShapiroTeukolsky}
\begin{equation}
\varepsilon_\nu^{nm}=\left(1.2\times 10^4\, {\rm erg~cm}^{-3}{\rm s}^{-1}
\right) \left(\frac{n}{n_0}\right)^{2/3}
\left(\frac{T}{10^7~{\rm K}}\right)^8\ .
\end{equation}
In evaluating the nuclear matter and quark matter
emissivities, we shall assume that the entire interior
of the star is at a common temperature $T$. Both nuclear
matter and quark matter are good conductors of heat,
and neutron stars older than a few years are well
approximated as isothermal.  

Because $\varepsilon_\nu \sim T^{5.5}$ in the gCFL phase, $L_\nu$
will be dominated by neutrino emission from the gCFL matter, unless
the gCFL volume is very small.  We include cooling curves
for cores made of ``unpaired quark matter'' even though 
this is not expected to be present on 
the phase diagram of QCD at neutron star temperatures
because it serves as a representative example
of the large class of phases of dense matter in which
$\varepsilon_\nu\sim T^6$
and $c_V\sim T$. This class includes all quark and nuclear
phases that cool by direct Urca processes, except for gCFL.

The surface of real neutron stars is colder than their
interiors, with the temperature gradients occurring only
in the outer envelope of the star within of order 100 meters
of the surface.  The heat transport within this envelope
has been analyzed~\cite{Gundmundsson}, and the result is well approximated by
a phenomenological relationship between the interior
temperature $T$ and the surface 
temperature $T_{\rm surface}$
given by~\cite{Gundmundsson,Page:2004fy}
\begin{equation}
T_{\rm surface}=(0.87\times 10^6~{\rm K})
\left(\frac{g_s}{10^{14} {\rm cm}/{\rm s}^2}\right)^{1/4}
\left(\frac{T}{10^8 {\rm K}}\right)^{0.55}\ ,
\label{surfaceinteriorrelation}
\end{equation}
where $g_s\equiv G_N M/R^2$ is the surface gravity.
This means that the rate of heat loss from the
surface of the star, which for a black body is 
\begin{equation}
L_\gamma=4\pi R^2 \sigma T_{\rm surface}^4 
\end{equation}
with $\sigma$ the Stefan-Boltzmann constant, is
given by
\begin{equation}
L_\gamma=4\pi R^2 \sigma (0.87\times 10^6~{\rm K})^4 
\left(\frac{g_s}{10^{14} {\rm cm}/{\rm s}^2}\right)
\left(\frac{T}{10^8 {\rm K}}\right)^{2.2}\ .
\end{equation}
We shall use this expression for $L_\gamma$, even though
we are not treating other aspects of the problem realistically,
because the fact that $L_\gamma\sim T^{2.2}$ will play
an important qualitative role.

The cooling of our ``star'' is described by the 
differential equation
\begin{equation}
\frac{dT}{dt} = - \frac{ L_\nu+L_\gamma}{V_{nm}c_V^{nm} + V_{qm}c_V^{qm}}
= - \frac{V_{nm}\varepsilon_\nu^{nm} + V_{qm}\varepsilon_\nu^{qm} + L_\gamma}
{V_{nm}c_V^{nm} + V_{qm}c_V^{qm}}
\label{coolingequation}
\end{equation}
which equates the heat lost ($\sim L dt$) to the  
change in the heat energy of the star ($\sim - V c_V  dT$).
We have all the ingredients needed to 
evaluate the right hand side of this equation in
place, with the exception of the specific heat of nuclear
matter and of unpaired quark matter. For a gas 
of several species of nointeracting
fermions, the specific heat is given by
\begin{equation}
c_V=\frac{k_B^2 T}{3\hbar^3 c}\sum_i 
p_F^i\sqrt{m_i^2 c^2 +(p_F^i)^2} \ ,
\end{equation}
where the sum runs over all the species.
In the case of noninteracting nuclear matter, the sum 
runs over $i=n,p,e$ and the Fermi momenta
for neutral matter in weak equilibrium are given by~\cite{ShapiroTeukolsky}
\begin{eqnarray}
p_F^n&=& \left(340~\MeV\right)\left(\frac{n}{n_0}\right)^{1/3}\nonumber\\
p_F^p&=&p_f^e = \left(60~\MeV\right)\left(\frac{n}{n_0}\right)^{2/3}\ .
\end{eqnarray}
We are taking noninteracting nuclear matter with density $n=1.5 n_0$ for the
mantle of our ``stars''.  In the case of neutral unpaired quark matter
in weak equilibrium, the sum on $i$ runs over the nine quarks and the Fermi
momenta are independent of color and are given by~\cite{Alford:2002kj}
\begin{eqnarray}
p_F^d &=&  \mu+\frac{M_s^2}{12\mu}\nonumber\\
p_F^u &=&  \mu-\frac{M_s^2}{6\mu}\nonumber\\
p_F^s &=&  \mu-\frac{5 M_s^2}{12\mu}\ ,
\end{eqnarray}
up to corrections of order $M_s^4/\mu^3$. We are using matter
with $\mu=500~\MeV$ in the core of our ``stars''.

\begin{figure}[t]
\begin{center}
\includegraphics[width=0.49\textwidth]{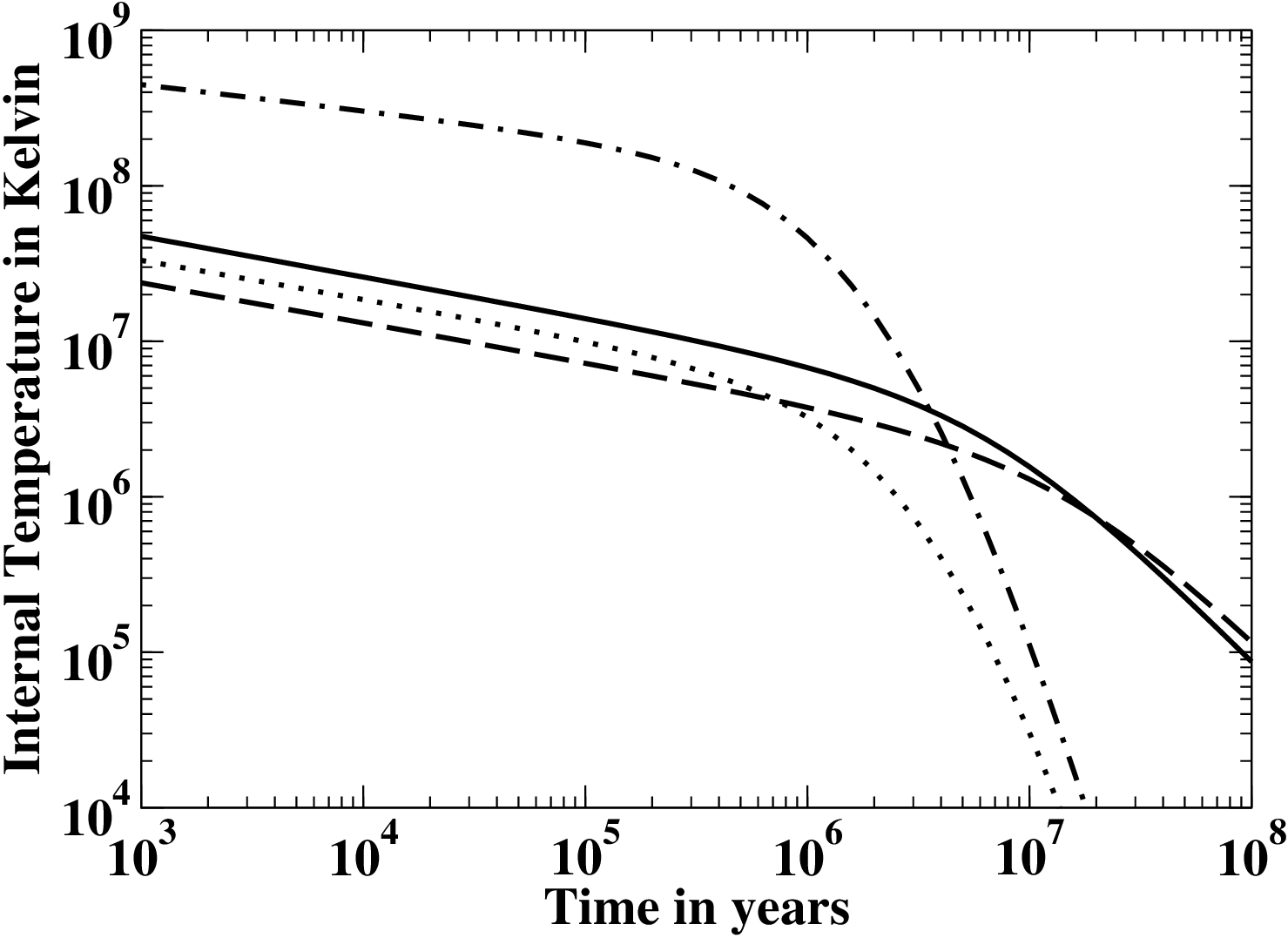}
\end{center}
\vspace{-0.25in}
\caption{Solutions to the cooling equation (\ref{coolingequation})
for 1.4 solar mass
``toy stars'' (described in the text) of four different compositions.
The curves show internal temperature as a function of time.
The dot-dashed curve is for a star with radius $R=12$~km
made entirely of nuclear matter
with a density $1.5 n_0$,
with no quark matter core. The other three curves describe
stars with radii $R=10$~km that have quark matter
cores with radii $R_{\rm core}=5$~km.
For all three curves, the quark matter has $\mu=500~\MeV$
and $M_s^2/\mu=100~\MeV$, with densities $\simeq 9 n_0$. 
For the dotted curve, the quark matter is noninteracting.
For the solid (dashed) 
curve, it is in the gCFL phase with $\De_0=25~\MeV$
($\De_0=40~\MeV$).
}
\label{fig:plot3}
\end{figure}

\begin{figure}[t]
\begin{center}
\includegraphics[width=0.49\textwidth]{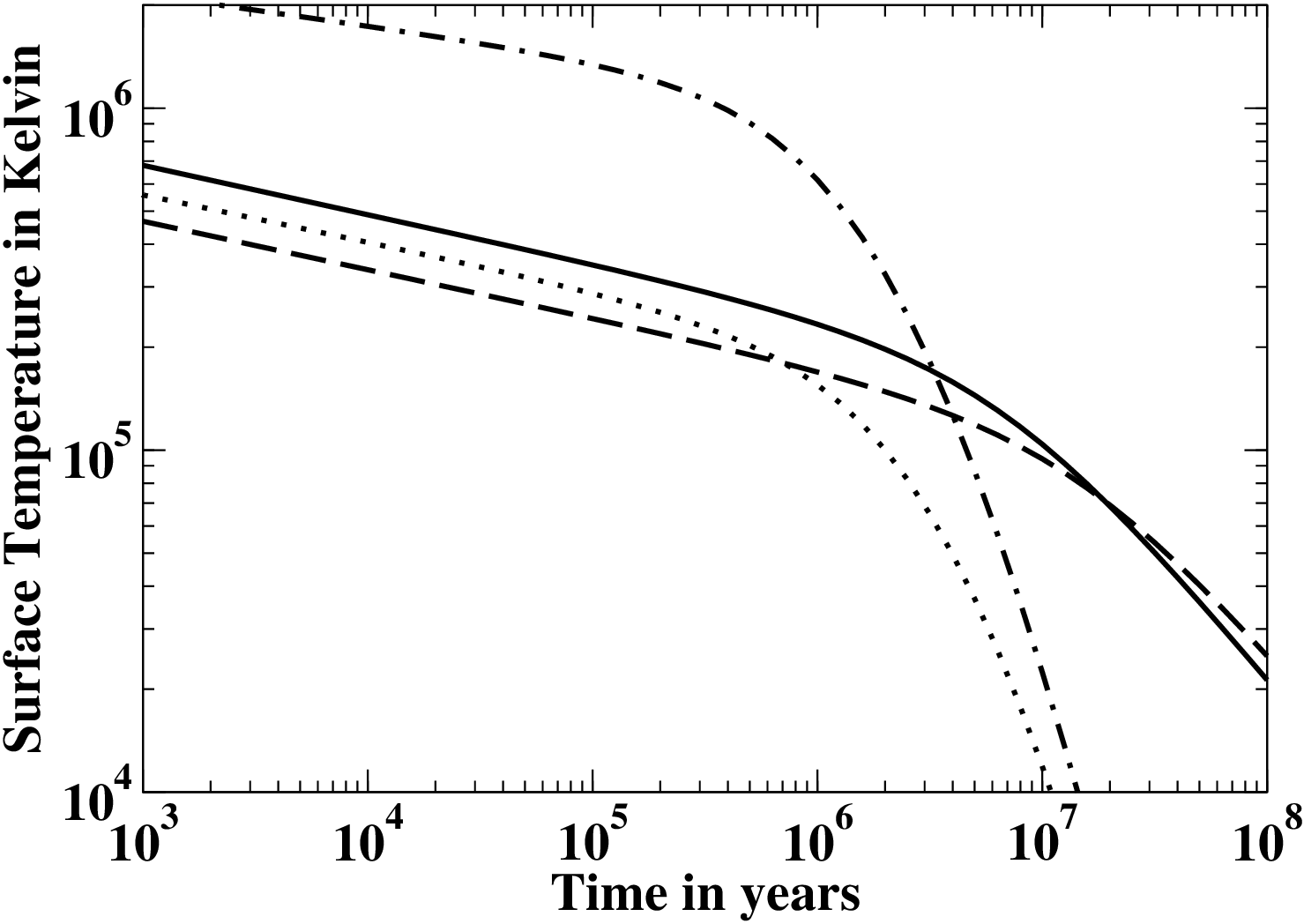}
\end{center}
\vspace{0.25in}
\caption{Same as Fig.~\ref{fig:plot3}, except that here we
plot $T_{\rm surface}$, related to the interior
temperatures plotted in Fig.~\ref{fig:plot3}
by Eq.~(\ref{surfaceinteriorrelation}).}
\label{fig:plot4}
\end{figure}

Fig.~\ref{fig:plot3} shows the cooling curves obtained
by solving (\ref{coolingequation}) for the four toy stars
we have described, plotted on
a log-log plot.  Each curve has an early time power law during
the period when cooling by neutrino emission dominates, namely
the first $10^{5}$ or so years.  At early times, $L_\gamma \ll L_\nu$
because $L_\gamma \sim R^2$ whereas $L_\nu\sim R^3$.  Because
$L_\nu$ drops much more rapidly than $L_\gamma$
as $T$ decreases, at late times $L_\gamma$ dominates and a new
power law is seen.  

It is easy to see why power law solutions arise.
In any temperature
regime in which the numerator and the denominator of the right hand
side of (\ref{coolingequation}) are each dominated by one
of their component terms, the cooling equation takes the form
\begin{equation}
\frac{dT}{dt} = - a T^p
\label{powerlawDE}
\end{equation}
for some $p$ and $a$.  For example, for a star that
is made entirely of nuclear matter, during
the epoch when $L_\nu \gg L_\gamma$ we have $p=7$, coming
from $L_\nu\sim T^8$ and $c_V^{nm}\sim T$.  For $p>1$, 
(\ref{powerlawDE}) has a power-law solution 
\begin{equation}
T=\left[ a(p-1) \, t \right]^{-\frac{1}{p-1}}\ .
\label{powerlaw}
\end{equation}
There are no arbitrary constants in this solution. 
We initialize the differential equation with some
temperature $T_0$ at a time $t_0=1$~year, chosen
because by that time the interior star can reasonably be treated
as isothermal. The initial condition $T_0(t_0)$ does not
appear in the power law solution: it only affects how
the power law solution is reached, if $T_0(t_0)$ does not lie on it.
Once the power law solution is reached, the form of the
solution to the
differential equation is independent of the initial condition.
We begin all our plots at $t=1000$~years, by which time the solution
is on the power law (\ref{powerlaw}) for any reasonable choice
of $T_0(t_0)$.  

During the epoch when $L_\nu\gg L_\gamma$,
a star made entirely of nuclear matter has $p=7$ and 
$T\sim t^{-1/6}$.  For the stars with unpaired quark matter, 
or gCFL quark matter, $p=5$ and $T\sim t^{-1/4}$ during
this epoch.  This explains how similar the three quark matter
core curves are during the first $10^5$ years,
and why all three stars with quark matter cores
are colder than the nuclear matter star.  Note that
the gCFL quark matter has $L_\nu\sim T^{5.5}$
and $c_V\sim T^{0.5}$, both enhanced by $1/T^{0.5}$ relative
to that of unpaired quark matter, and indeed relative to any 
phase of nuclear or quark matter in which direct Urca
processes occur that has been considered previously.
But, the effect of these enhancements cancel in the cooling
curve during the epoch when $L_\nu \gg L_\gamma$.  
There are now a number of 
indications~\cite{Slane:2002ta} 
that some
neutron stars with ages of order $10^3$ to $10^{5}$ years
(presumably the heavier ones, although this is certainly
not demonstrated)
are significantly colder than would be expected in the absence
of direct Urca neutrino emission, whereas other (presumably less
massive) stars have temperatures consistent with 
theoretical cooling curves calculated upon assuming
nuclear matter composition.  Were this to be confirmed, the
discovery of direct Urca emission with $T\sim t^{-1/4}$,
instead of the slower $T\sim t^{-1/6}$, could indicate 
the presence of any number of
dense matter phases, including gCFL quark matter but
also including nuclear matter leavened with either hyperons, kaons
or pions.  

At late times, when $L_\gamma\gg L_\nu$ all stars except those
containing gCFL quark matter
have $p=2.2-1=1.2$ because $L_\gamma\sim T^{2.2}$ and $c_V\sim T$,
and hence cool with $T\sim t^{-1/0.2}=t^{-5}$.
This explains
the rapidly dropping temperatures at late times
for the stars without
gCFL quark matter in Fig.~\ref{fig:plot3}.  
If the volume of gCFL matter is sufficient (more
on this below) it dominates the heat capacity
of the star, yielding $p=2.2-0.5=1.7$
because $c_V\sim T^{0.5}$,
and hence the star cools
with $T\sim t^{-1/0.7} = t^{-1.4}$ at late times.  
The gCFL
matter keeps the aging star warm by virtue of its large heat
capacity.  Hence, the title of our paper.

We show the surface temperatures of our toy stars in Fig.~\ref{fig:plot4}.
It is tempting to put data obtained from the observation of
real stars on this plot, but we resist the temptation given
that our ``stars'' are not realistic.  The qualitative
impact of gCFL quark matter is, however, clear: stars which
are old enough that they cool by photon emission stay
much warmer if they contain a gCFL hot water bottle.
In our Fig.~\ref{fig:plot4}, which
should be taken as illustrative and not
yet as a quantitative prediction, the effect is a full order
of magnitude for $10^7$-year old stars, and gets much larger
for older stars, as the cooling curves of all stars except
those containing gCFL quark matter drop rapidly.

\begin{figure}[t]
\begin{center}
\includegraphics[width=0.49\textwidth]{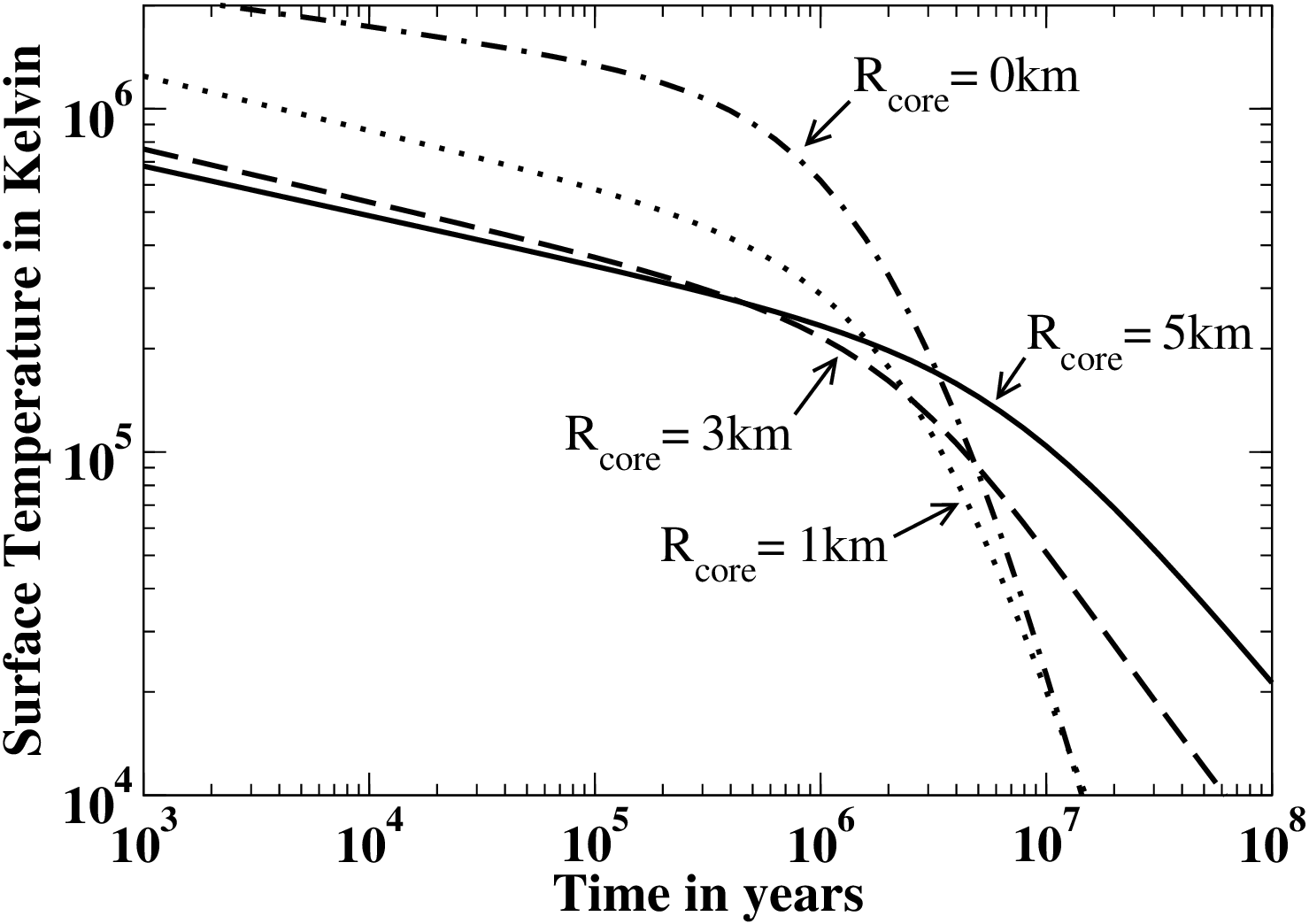}
\end{center}
\vspace{-0.25in}
\caption{Cooling curves showing the surface temperature
of stars with gCFL cores with $\De_0=25~\MeV$ for 
$R_{\rm core}=5$~km (solid; same as solid curve in Fig.~\ref{fig:plot4}),
$R_{\rm core}=3$~km (dashed),  $R_{\rm core}=1$~km (dotted),
$R_{\rm core}=0$~km (dot-dashed; same as dot-dashed curve in 
Fig.~\ref{fig:plot4}).}
\label{fig:plot5}
\end{figure}

In Fig.~\ref{fig:plot5}, we investigate the dependence of the cooling
curves on the volume of gCFL quark matter present  in the core
of the star. We  see that the ``hot water bottle effect'' is
present for $R_{\rm core}=3$~km, but reduced in magnitude.
For $R_{\rm core}=1$~km, no effect is visible: the effect does
occur, but only at even lower temperatures than we have plotted.
(Because its heat capacity is $c_V\sim T^{0.5}$, if
any gCFL quark matter is present it will eventually dominate
the heat capacity of the entire star, no matter how
small its volume fraction.  For $R_{\rm core}=1$~km,
this occurs at temperatures below those we have plotted.)
Note that what we are referring to as $R_{\rm core}=5$~km
could equally well describe a star with a shell of gCFL
quark matter extending between radii of 4.5 and 6 km.

\section{Outlook}
\label{sec:outlook}

We hope that our results challenge observers to 
constrain the temperature of neutron
stars that are 10 million years old or older.  Prior to
our work, all proposed cooling curves for these old  
stars drop so fast into unobservability that there has been 
little motivation to make the effort to obtain the best
constraints possible on their temperatures.  Given that
we know of isolated
neutron stars that are younger than a million years
old and closer than 200 parsecs, it is reasonable to expect
that there are $10^7$-year old isolated neutron stars
closer than 100 parsecs to earth.  At first we were concerned
that if such old nearby stars had temperatures of order $10^5$~K,
as Fig.~\ref{fig:plot4} suggests will be the case if they
contain gCFL hot water bottles, they should already have been
detected.  However, initial estimates suggest that
they will in fact be quite a challenge to find, since the
peak of a $10^5$~K black body spectrum lies in the far
ultraviolet, where the interstellar medium is opaque, and
since they will be quite faint in the accessible UV and visible
wavelengths~\cite{KaplanPrivate}. 
Another option, perhaps easier than finding these
stars without knowing where to look,
is to study nearby old pulsars,
already detected by their nonthermal emission, and to constrain
their thermal emission hence putting an upper bound on their 
temperature. This has been done for PSR 0950+08, whose
spin-down age is $10^{7.2}$ years, 
yielding the bound $T<10^{5.2}~{\rm K}$~\cite{Zavlin:2004wt}.  
This limit is quite promising, as it is close to the 
curves in Fig.~\ref{fig:plot4} describing the cooling of our toy
star with a gCFL core. 
And,
we are confident
that we have not thought of the best way of looking for
aging but still warm neutron stars. We are 
therefore hopeful that the
opportunity to make an unambiguous discovery of the 
presence of quark matter within neutron stars or
to rule out the presence of gCFL quark matter in the
entire region of the QCD phase diagram sampled by neutron stars
will stimulate observers to rise to the challenge.

Much theoretical work remains to be done. 
Interesting microphysical questions
about the gCFL phase remain, and have been enumerated
in Refs.~\cite{Alford:2004hz,Fukushima:2004zq}.  
Phases with some features in common with the gCFL
phase can be unstable with respect to inhomogeneous
mixed phases~\cite{Bedaque:2003hi}, and although the gCFL phase
is stable with respect to all straightforward mixed
phase possibilities~\cite{Alford:2004hz,Alford:2004nf}, 
an exhaustive investigation has not yet been performed.
Perhaps the most interesting open questions
are the possibilities of 
$K^0$-condensation~\cite{Kryjevski:2004jw,Forbes:2004ww}
or gluon condensation~\cite{Huang:2004bg} in the gCFL phase.
Either could change our quantitative results
for its $c_V$ and $\varepsilon_\nu$, but neither is likely to change
their unusual $T$-dependence: $c_V\sim T^{0.5}$ and
$\varepsilon_\nu\sim T^{5.5}$. (Neither $K^0$-mesons 
nor gluons~\cite{Alford:2002kj,Gerhold:2004sk} 
would affect the $\tilde Q$-charge balance, which
is responsible for the existence of the gapless quasiparticle
with a quadratic dispersion relation in Fig.~\ref{fig:disprel}, 
whose consequence in turn
is the unusual $T$-dependence of the gCFL $c_V$ and $\varepsilon_\nu$.)
As far as theoretical astrophysical 
work, our results for $c_V$ and $\varepsilon_\nu$
must be incorporated into calculations of cooling curves for
stars with realistic atmospheres and density
profiles before plots like those in Figs.~\ref{fig:plot4}
and \ref{fig:plot5} can be compared quantitatively to data.
Nonetheless, our conclusion that
gCFL quark matter within a neutron star  
will keep the star warm in its old age relies
only on the unusual $T$-dependence of the gCFL specific heat,
and is therefore expected to be robust.

\begin{acknowledgments}

We thank David L. Kaplan for orientation on various
questions related to observations and Kenji
Fukushima for assistance with quasiparticle
dispersion relations, and 
acknowledge helpful conversations with 
them and with Michael Forbes. We thank Igor Shovkovy for
finding an error in a previous version of this paper.
PJ is grateful to the Research Science Institute of
the Center for Excellence in Education for supporting
her research.
This research was supported in part by DOE grants 
DE-FG02-91ER40628, DF-FC02-94ER40818, and DE-FG02-93ER40762.
\end{acknowledgments}

\end{document}